\documentclass[journal]{IEEEtran}
\usepackage{algpseudocode}
\usepackage{amsmath,amssymb,amsfonts}
\usepackage{amsthm}
\usepackage{algpseudocode}
\usepackage{algorithm}
\usepackage{mathrsfs}
\usepackage{xcolor}

\def\BibTeX{{\rm B\kern-.05em{\sc i\kern-.025em b}\kern-.08em
    T\kern-.1667em\lower.7ex\hbox{E}\kern-.125emX}}
\usepackage{array}
\usepackage[caption=false,font=normalsize,labelfont=sf,textfont=sf]{subfig}
\usepackage{textcomp}
\usepackage{stfloats}
\usepackage{url}
\usepackage{verbatim}
\usepackage{graphicx}
\usepackage{cite}
\newtheorem{proposition}{Proposition}
\usepackage{multirow}

\ifCLASSINFOpdf

\else

\fi

\begin{document}
%
\title{Adaptive Semantic Communication for UAV/UGV Cooperative Path Planning}
%
%
%

\author{Fangzhou Zhao,~\IEEEmembership{Graduate Student Member}, Yao Sun$^{\star}$,~\IEEEmembership{Senior Member,~IEEE}, Jianglin Lan, Lan Zhang,~\IEEEmembership{Member,~IEEE}, Xuesong Liu,~\IEEEmembership{Graduate Student Member}, and Muhammad Ali Imran,~\IEEEmembership{Fellow,~IEEE}
\thanks{Fangzhou Zhao, Yao Sun, Jianglin Lan, Xuesong Liu and Muhammad Ali Imran are with the James Watt School of Engineering, University of Glasgow, Glasgow G12 8QQ, U.K.}
\thanks{*Jianglin Lan was funded by the Leverhulme Trust Early Career Fellowship (ECF-2021-517).}
\thanks{Lan Zhang is with Department of Electrical and Computer Engineering, Clemson University, Clemson, South Carolina 29634, USA (e-mail: lan7@clemson.edu).}
\thanks{$^\star$Corresponding author. (Email: Yao.Sun@glasgow.ac.uk)
}}

\maketitle

\begin{abstract}
Effective path planning is fundamental to the coordination of unmanned aerial vehicles (UAVs) and unmanned ground vehicles (UGVs) systems, particularly in applications such as surveillance, navigation, and emergency response. Combining UAVs' broad field of view with UGVs' ground-level operational capability greatly improve the likelihood of successfully achieving task objectives—such as locating victims, monitoring target areas, or navigating hazardous terrain. In complex environments, UAVs need to provide precise environmental perception information for UGVs to optimize their routing policy. However, due to severe interference and non-line-of-sight conditions, wireless communication is often unstable in such complex environments, making it difficult to support timely and accurate path planning for UAV-UGV coordination.
To this end, this paper proposes a semantic communication (SemCom) framework to enhance UAV/UGV cooperative path planning under unreliable wireless conditions. Unlike traditional methods that transmit raw data, SemCom transmits only the key information for path planning, reducing transmission volume without sacrificing accuracy. The proposed framework is developed by defining key semantics for path planning and designing a transceiver for meeting the requirements of UAV-UGV cooperative path planning. Simulation results show that, compared to conventional SemCom transceivers, the proposed transceiver significantly reduces data transmission volume while maintaining path planning accuracy, thereby enhancing system collaboration efficiency.
\end{abstract}


%
\IEEEpeerreviewmaketitle

\section{Introduction}
In recent years, the significant advancements in the autonomy, intelligence, and environmental adaptability of unmanned systems have led to the widespread deployment of unmanned aerial vehicles (UAVs) and unmanned ground vehicles (UGVs) in increasingly complex scenarios, such as surveillance, navigation, crowd control, sensing, and emergency response\cite{UAV3,UAV11,UAV22}. By leveraging the collaboration between UAVs and UGVs, tasks requiring both aerial surveillance and ground-level operations, such as search and rescue missions, environmental monitoring, and infrastructure inspection, can be effectively addressed. UAVs, with their broad field of view and agile maneuverability, provide valuable perception data, while UGVs not only execute specific tasks such as target tracking, environmental exploration, or cargo transportation, but also possess comparatively rich computational and energy resources, thereby enabling a comprehensive and efficient solution for complex applications\cite{UGV1, UAV-M}.

In UAV/UGV cooperative control tasks, path planning plays a fundamental role in determining the overall efficiency, responsiveness, and safety of mission execution \cite{UGV2,UAV44}. Path planning enables both aerial and ground platforms to coordinate their trajectories, avoid obstacles, and reach target locations while optimizing resource utilization, such as energy and bandwidth. However, a primary challenge in such systems is the distributed decision-making architecture, where UAVs and UGVs rely on continuous information exchange—such as position updates, sensor data, and planning-related commands—to coordinate their traversal routes effectively \cite{UGV2,UAV11}.

The performance of the communication system is a critical factor that directly impacts the success of path planning, especially in dynamic and uncertain environment. In practical scenarios such as post-disaster search and rescue missions, underground tunnels, or dense urban areas, wireless links often suffer from harsh channel conditions, including interference, signal blockage, multipath fading, and limited spectrum resources. These issues can cause significant communication degradation, leading to intolerable latency, packet loss, or even temporary disconnection \cite{UAV-COM,UAV33}.
Such impairments severely limit the timely exchange of critical control and perception information, preventing UAVs and UGVs from updating their paths based on the latest environmental feedback. As a result, UGVs may operate with inaccurate environmental information, leading to suboptimal routing decisions and reduced mission effectiveness. Therefore, ensuring robustness is essential for supporting reliable and adaptive path planning in UAV/UGV cooperative systems.

To address the aforementioned communication challenges, it is essential to design an efficient and reliable communication strategy to ensure that UAVs and UGVs can stably share mission-critical information in complex environments. However, developing such a communication strategy presents several challenges. Since UAVs have limited onboard resources, it is important to design resource-efficient communication strategies to avoid unnecessary computational overhead. Hence, the system requires a lightweight and reliable communication method with low computational cost. Moreover, the complex channel variations and interference in dynamic environments demand that encoding and decoding strategies rapidly adapt to changes to ensure the accurate transmission of critical information.

Semantic communication (SemCom) offers an opportunity to address the aforementioned communication challenges. Unlike traditional communication systems that transmit all bits of raw data, SemCom aims to convey the intended meaning of the information, focusing on the semantics rather than the bits \cite{chengsi,zhao2024enhancing}. This enables the communication process to prioritize task-relevant content over bit-level accuracy.
Recent studies have explored the application of SemCom in autonomous systems, demonstrating its potential to improve communication efficiency and decision-making in UAV and UGV scenarios \cite{uav, u2}.
This approach can reduce data transmission volume and enhances communication efficiency \cite{scc}. Moreover, by leveraging deep learning techniques, SemCom has the potential to adapt transmission strategies based on task demands and communication conditions \cite{sc2, sc1}. By reducing bandwidth consumption whilst improving the reliability of collaborative path planning, SemCom provides a better communication solution for UAV/UGV cooperative path planning. However, directly exploiting traditional SemCom for UAV/UGV cooperation is not applicable. 
On resource-constrained UAV platforms, SemCom should be lightweight to ensure deployability, but excessive simplification can compromise the reliability of transmission.
To address this, it is necessary to develop a tailored SemCom transceiver that focuses on the path-planning key semantics while balancing efficiency and reliability in UAV/UGV cooperation.

In this paper, we propose a control‑aware SemCom framework for the UAV/UGV cooperative path planning, and instantiate it with a dynamically sparsifiable transceiver guided by path planning-critical semantics. We mathematically prove the conditions of maximizing path planning accuracy and design a transceiver named Path-SC that dynamically adjusts the SemCom encoding/decoding strategy base on the path planning-critical semantics.
The main contributions of this paper are summarized as follows:
\begin{itemize}
\item 
\textit{A SemCom framework for UAV/UGV cooperative path planning:}
We address a resource-constrained, image-based wireless scenario involving cooperative path planning between UAVs and a UGV, by proposing a SemCom framework for efficient transmission of path planning-critical information.
In this framework, multiple UAVs capture ground images and transmit them to the UGV via wireless SemCom. The UGV then utilizes the received semantic information to construct a map and perform path planning. The degree of sparsification is dynamically adjusted based on the importance of the regions in the map. Crucial areas for path planning, such as obstacle boundaries and traversable regions, are assigned a higher degree of sparsification, ensuring accurate transmission of essential features, while less important areas are sparsified to reduce transmission load.

\item
\textit{Semantic identification method for path planning-critical regions:}
We mathematically derive the conditions necessary for identifying the image regions containing the semantic information critical to path planning. Building on these conditions, we develop a method to accurately locate these path planning-critical regions and use them to guide the transmission strategy for each UAV, ensuring that more computing resources are devoted to semantically important areas. This targeted allocation directly improves the reliability of semantic information delivery and enhances communication efficiency in dynamic environments.

\item
\textit{Adaptive SemCom transceiver for path planning:}
We develop a novel SemCom transceiver, named \textit{Path-SC}, equipped with a dynamically sparsifiable encoder. The Path-SC applies a more robust and reliable encoding to the identified critical image regions for path planning, while sparsifying less significant areas to reduce computational complexity and improve transmission efficiency. This adaptive strategy ensures effective resource utilization while maintaining the accuracy of path planning critical information.

\end{itemize}

The rest of this paper is organized as follows. Section II provides a review of the related work. Section III presents the SemCom framework for UAV/UGV cooperative path planning. Section IV describes the method for identifying the semantics of path planning-critical regions. Section V details the SemCom transceiver design. Section VI reports the numerical simulations. Section VII concludes the paper.

\section{Related Work}
With the complexity of UAV/UGV cooperative missions, ensuring efficient control, communication, and computation is a critical challenge. Recent research has made significant progress in both SemCom and collaborative autonomous systems communication. This section reviews advances in communication for cooperative autonomous systems and development in SemCom for task-oriented wireless systems.

\subsection{Wireless Communication in Cooperative UAV/UGV Systems}
Recent advancements in UAV/UGV cooperative systems have led to significant progress in path planning, obstacle recognition, and task execution. Various studies have explored innovative solutions to address these challenges, demonstrating effective approaches for improving operational efficiency and adaptability in complex environments. 

For instance, the work of \cite{UGV2} proposes an automatic ground map building and hybrid path planning algorithm that utilizes UAV-captured images for obstacle recognition. By employing a combination of genetic algorithm and local rolling optimization, it achieves real-time, cost-efficient path planning. The authors in \cite{UGV1} focus focuses on illegal urban building detection, introducing a two-level memetic algorithm to minimize overall execution time. By considering factors such as UGV speed, UAV power limits, and communication constraints, the proposed method demonstrates its superiority over state-of-the-art approaches. The study of \cite{UGV3} investigates a UAV-oriented computation offloading system within an air-ground cooperative network. Using a primal decomposition approach, it jointly optimizes UAV energy efficiency by accounting for mobility, air-to-ground communication, and computation dynamics, achieving notable performance gains compared to benchmark methods. Additionally, the study of \cite{ugv4} proposes a secure communication strategy for offloading computation-intensive tasks from UAVs to UGVs, addressing issues like eavesdropping and UGVs leaving the target area. A non-convex mixed integer nonlinear programming problem is solved using an iterative algorithm that combines block coordinate descent and successive convex approximation to efficiently optimize latency, power, velocity, anti-collision, and distance.

While these studies have made considerable progress in UAV/UGV cooperative path planning, the collaborative nature of these systems imposes significant communication challenges. The need for timely and reliable information exchange is critical for coordinated decision-making and path adjustments. However, dynamic environments and resource constraints often hinder effective communication. Therefore, recent studies such as \cite{oubbati2025uav,dutriez2024energy} have focused on optimizing energy management in UAV/UGV systems through adaptive energy-efficient communication strategies, aiming to enhance both communication and energy efficiency in cooperative tasks.
Some research, such as \cite{alotaibi2025optimizing}, focuses on optimizing computational resource overheads in resource-constrained UAV platforms. These studies lay the groundwork for enhancing UAV mission performance by deploying advanced communication modules. 
However, while these methods help in the deployment of more sophisticated communication systems on UAVs, they do not tackle the particular communication needs in path planning. In path planning, the key requirement is transmitting semantic details of the environment—like obstacles and road structure—because inaccuracies here directly affect the correctness of the planned path.

\subsection{Advances in Wireless SemCom Systems}
SemCom offers a promising opportunity to tackle these challenges by focusing on transmitting task-relevant information instead of raw data, thereby improving communication efficiency and reliability in UAV/UGV cooperative control tasks. 

For example, the research in \cite{sc2} proposes a semantic coding method utilizing a generative adversarial network (GAN) to effectively reconstruct key features from the transmitted data. This approach ensures the preservation of essential semantics, even under poor bit rates, by enhancing the network's ability to recover crucial information that might otherwise be lost due to bandwidth limitations. The research in\cite{sc1} presents a robust SemCom framework that enhances noise resilience through adversarial training. By incorporating semantic noise samples into the training dataset, the framework improves the transceiver's robustness, ensuring more reliable performance in noisy environments. The work of \cite{runze} applies SemCom to AI-generated content tasks, integrating a diffusion model into the semantic encoder and decoder. This approach efficiently transmits intermediate features, enhancing the quality and efficiency of content transmission. The work of \cite{sc3} proposes a semantic-aware radio access network to address the limitations of existing SemCom research in real-world radio access networks. 

Despite these advancements, existing SemCom methods are not directly applicable to UAV/UGV cooperative path planning. Most current approaches are designed for general multimedia transmission, neglecting the dynamic and control-sensitive nature of UAV/UGV tasks. Therefore, this paper proposes a SemCom framework tailored for UAV/UGV cooperative path planning, ensuring the reliable transmission of task-critical information while optimizing communication efficiency in complex environments.

\begin{figure*}[htp]
	\centering
	\includegraphics[width=0.9\textwidth]{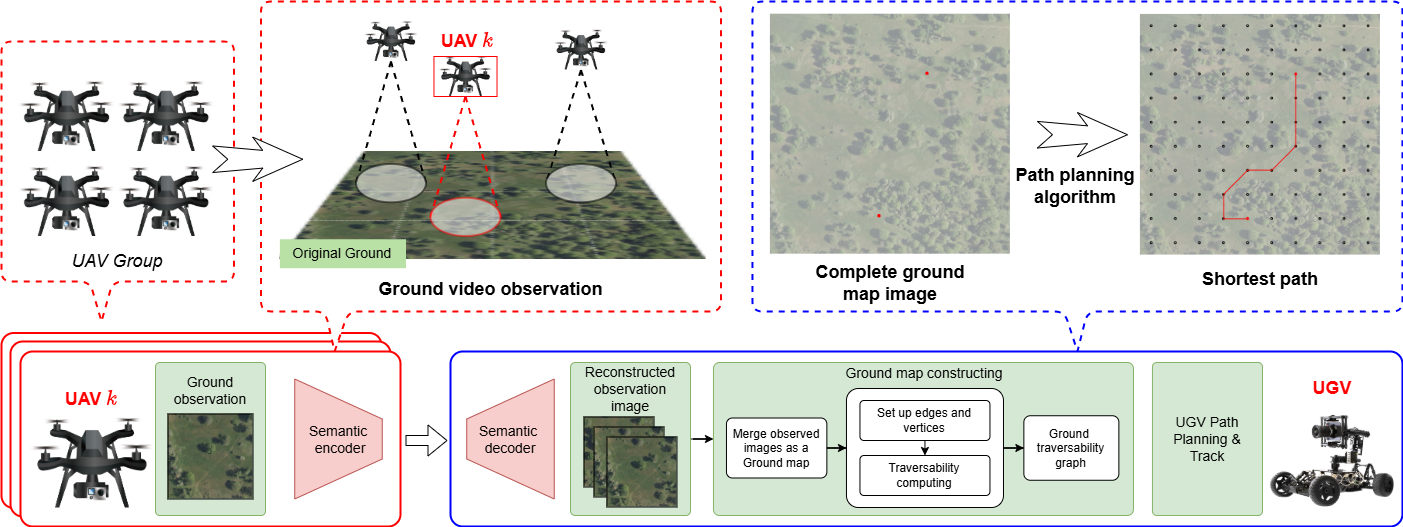}
	\caption{The proposed SemCom framework for UAV/UGV cooperative path planning.}
	\label{img1}
\end{figure*}

\section{SemCom Framework for UAV/UGV Cooperative Path Planning}
Consider a UAV/UGV cooperative system comprising multiple UAVs and a single UGV. As shown in Fig. 1, the UAV first conducts ground observations and transmits the captured image to the UGV via SemCom. Upon receiving the observation data, the UGV processes the image to find the target position and construct a ground map. Based on its location, the target position, and the ground map, the UGV executes the path planning algorithm to obtain a feasible route to the destination and simultaneously determines the corresponding communication strategy for semantic transmission along the planned path. In this system, only the semantic encoder runs on the UAV, while the decoder and all map construction and path-planning modules execute on the UGV, so that the heavier workloads are handled by the UGV and the on-board UAV computation is limited to the encoder’s forward pass.

\subsection{UGV Path Planning}
To perform path planning, the UGV first constructs a ground map based on the image collected by the UAV.
At time $t$, the images collected for the UGVare denoted as $\mathcal X_t=\{\mathbf x_{t}^1, \mathbf x_{t}^2, \cdots, \mathbf x_{t}^n\}$, where $\mathbf x_{t}^{k}$ represents the image collected by the $k$th UAV. 
These images are sent to the UGV and combined into a complete ground map image $\mathbf X_t$. 
However, the ground image taken by the UAV is inevitably subject to noise during transmission, leading to distortion of ground features. As a result, the ground image received by the UGV deviates from the original $\mathbf X_t$. We denote the received ground images set as $\mathcal {\hat X}_t=\{\mathbf {\hat x}_{t}^{1}, \mathbf {\hat x}_{t}^{2}, \cdots, \mathbf {\hat x}_{t}^{n}\}$ and the received ground image as $\mathbf {\hat X_t}$.

For the received $\mathbf {\hat X_t}$, we analyze the ground traversability of the UGV and generate a traversability graph for planning the path \cite{trav}. 
The traversability $\tau$, taking values within the interval $[0,1]$, measures the difficulty for the UGV to navigate a given area. For example, the flat areas free of obstacles exhibit a higher traversability, while regions characterized by steep slopes, obstacles, or cliffs demonstrate a reduced traversability. 
The traversability graph is constructed by gridding the map, where each grid is treated as a vertex with weight $\tau$ and neighboring vertices are connected by weighted edges in horizontal, vertical, and diagonal directions. The traversability graph of $\mathbf X_t$ is represented as a matrix of the vertex weights, denoted by $\mathbf G_t$. The weight of the edge between the vertices $u$ and $v$ is given by 
\begin{equation}
    w\left(u,v\right)= \left(\frac{\kappa}{\tau_{u}} \right)^2 + \left(\frac{\kappa}{\tau_{v}} \right)^2,
\end{equation}
where $\kappa$ is the coefficient, and $\tau_{u}$ and $\tau_{v}$ are the weights of $u$ and $v$, respectively.

Due to the distortion of the transmitted ground image, the weights in the traversability graph of the received ground image $\mathbf {\hat X}_t$ are perturbed. We denote the traversability graph of $\mathbf {\hat X}_t$ as $\mathbf {\hat G}_t$. The relation between $\mathbf G_t$ and $\mathbf {\hat G}_t$ is described by 
\begin{equation}
    \mathbf {\hat G}_t=\mathbf G_t+\mathbf E_t,
\end{equation}
where $\mathbf E_t$ is a matrix of random variables that represent the weight perturbations after wireless transmission. For the vertex $u$, the weight perturbation is denoted as $\epsilon_u$.

Based on the traversability graph, any shortest path algorithm can be executed to determine the optimal route from the starting point to the target area. In this work, we use Dijkstra's algorithm, a well-known algorithm for computing shortest paths in weighted graphs. The shortest path computed by Dijkstra's algorithm based on the $\mathbf {\hat X}_t$ is denoted as $\mathbf {\hat p}_t^*$, while the shortest path computed based on the original image $\mathbf X_t$ is denoted as $\mathbf p_t^*$, where $\mathbf {\hat p}_t^*$ and $\mathbf {p}_t$ are the ordered set of vertices.
It is worth emphasizing that our SemCom framework and the subsequent analysis are independent of the specific path planning algorithm.
The Dijkstra algorithm is adopted here as a standard shortest-path algorithm to isolate the effectiveness of semantic transmission.

\subsection{SemCom Model}
We exploit SemCom to transmit $\mathcal {X}_t$ from the UAV to UGV. 
In high-performance neural network-based SemCom, images are typically partitioned into multiple small patches for encoding and decoding, with each patch differing in semantic significance and contributing variably to the overall image comprehension \cite{sc1}.
Thus, the image $\mathbf x_{t}^{k}$ can be represented as $\mathbf x_{t}^{k}=\{x_{t}^{k1},x_{t}^{k2},\cdots,x_{t}^{km}\}$, where $x_{t}^{ki}$ denotes the $i$-th patch of the $\mathbf x_{t}^{k}$. 

Due to the rate and latency constraints in UGV navigation tasks, it is infeasible for each UAV transceiver to perform the highest-quality coding. Intuitively, priority is given to enhancing patches with higher semantic importance for UGV navigation to improve the transmission accuracy of $\mathbf x_{t}^{k}$ via SemCom.
In particular, in our framework, semantic importance refers to the density of task-relevant features within a visual patch, specifically those that contribute to path planning decisions, such as obstacle boundaries or traversable regions.
The patches with high semantic significance in $\mathbf x_{t}^{k}$ are prioritized for accurate transmission, while other patches can undergo simplified encoding to accommodate the limited transmission rate and further reduce computing time.

Meanwhile, the proportion of patches with high semantic importance in $k$th UAV is denoted as $\delta_k$. This parameter is dynamically adjusted based on the importance of different regions in the ground image. Specifically, regions that are crucial for path planning, such as obstacle boundaries and traversable regions, are assigned a higher \(\delta_k\) to ensure more accurate transmission of these semantic features. In contrast, less important regions, which do not significantly impact path planning, are assigned a lower \(\delta_k\), reducing the transmission load and computational cost. This dynamic adjustment helps the system balance communication efficiency with the accuracy required for effective path planning, adapting to the varying importance of image regions in different environments.
Intuitively, a small $\delta_k$ can reduce the number of bits being transmitted, thereby reducing the energy and the information transmission delay. However, this reduction could result in a compromise on image accuracy, increasing the discrepancy between $\mathbf {\hat X}_t$ and $\mathbf X_t$ in the UGV path planning system.



For the SemCom-based image transmission, $\mathbf x_{t}^{k}$ is first encoded into bits representing semantics $\mathbf s_{t}^{k}$ by a semantic encoder, as follows:
\begin{equation} 
\mathbf s_{t}^{k} = {\mathcal{ E}_e}\left ({{\mathbf {x}_{t}^{k}}; { \boldsymbol \alpha, \delta_k}} { }\right),
\end{equation}
where ${\mathcal{ E}_e}$ is the neural networks of the encoder with the trainable parameters $\boldsymbol \alpha$.

After $\mathbf s_{t}^{k}$ are transmitted through the noisy physical channel, the received bits $\mathbf {\hat s}_{t}^{k}$ are
\begin{equation} 
\mathbf {\hat s}_{t}^{k} = {\mathbf h(t) *\mathbf s_{t}^{k} + {\mathbf {n}}},
\end{equation}
where $\mathbf h(t)$ represents the channel coefficient, and $\mathbf {n}$ is the additive white Gaussian noise (AWGN) of the channel.
Correspondingly, the reconstructed image is finally represented as
\begin{equation} 
{\mathbf {\hat x}}_{t}^{k} = {\mathcal{ E}_d}\left ({\mathbf {\hat s}_{t}^{k}; { \boldsymbol \alpha, \delta_k}} {}\right),
\end{equation}
where ${\mathcal{ E}_d}$ is the neural networks of the semantic decoder.

\section{Transmission Strategy for Path Planning Critical Regions}
\subsection{Problem Formulation}
SemCom can be leveraged to reduce image distortion and improve the accuracy of traversability maps for path planning. However, due to constraints on transmission rate and computing resources, it is infeasible to transmit all the semantic information. Therefore, we need to design a method for optimizing the transmission strategy of each UAV to maximize the accuracy of path planning.

In the process of wireless transmission and semantic reconstruction, the distortion of the ground image comes from the combination of multiple errors, including the reception disturbance caused by physical channel noise and fading, the reconstruction error caused by semantic encoding and decoding, the quantization/compression error caused by bit budget constraint, and the perception error caused by occlusion and sensor imperfections. The resulting weight perturbation of the vertex $u$ is denoted as $\epsilon_u$.
Let $\mathbf{P}_t$ be the set of all possible paths between a given source and destination point on $\mathbf{G}_t$, and $\mathbf{p}_t^*$ be the shortest path in $\mathbf{P}_t$.
The total weight of a path $\mathbf{p}_t^*$ is denoted by $\mathbf{W}(\mathbf{p}_t^*)$, defined as the sum of vertices weights along the path. For a candidate path $\mathbf{p}_t^j \ne \mathbf{p}_t^*$, the weight difference between $\mathbf{p}_t^*$ and $\mathbf{p}_{t}^{j}$ is denoted as $\Delta w_j = \mathbf{W}(\mathbf{p}_t^*) - \mathbf{W}(\mathbf{p}_{t}^{j})$.

In particular, since $\hat{\mathbf{G}}_t$ shares the same topology as $\mathbf{G}_t$, both paths $\mathbf{p}_t^*$ and $\mathbf{p}_{t}^{j}$ should be the same on $\mathbf {\hat G}_t$ and $\mathbf {G}_t$, but with different weights. 
The weight difference of $\mathbf p_{t}^{j}$ and $\mathbf {p}_t^*$ on $\mathbf {\hat G}_t$ is denoted as
\begin{align}
    {\Delta {\hat w}_j}=\Delta w_j+\sum\limits_{u \in \mathbf {p}_t^*} {\epsilon_u}-\sum\limits_{u \in \mathbf p_{t}^{j}} {\epsilon_u}.
\end{align}
with
\begin{align}
\mathrm{Var}(\Delta\hat w_j) = \sigma_{p_t^*}^2 + \sigma_{p_t^j}^2.
\end{align}
where $\sigma_{\mathbf{p}_t^*}^2 = \sum_{u \in \mathbf{p}_t^*} \epsilon_u$ and $\sigma_{\mathbf{p}_{t}^{j}}^2 = \sum_{u \in \mathbf{p}_{t}^{j}} \epsilon_u$.
Please note that we model $\epsilon_u$ as Gaussian in the subsequent analysis. The Gaussian behavior can be theoretically justified by the delta method and the central limit theorem. Specifically, let $\hat{\tau}_u = g(x+e)$ be the estimated traversability of vertex $u$, where $x$ is the noise-free image patch and $e$ is the aggregated error vector composed of the aforementioned four types of distortions. A first-order Taylor expansion yields
\begin{align}
\epsilon_u = g(x+e) - g(x) \approx \nabla g(x)^\top e.
\end{align}
When the components of $e$ consist of multiple independent or weakly correlated small perturbations, their linear combination $\epsilon_u$ converges in distribution to a Gaussian.

Then in the Gaussian case, the probability that $\mathbf {p}_t^*$ is shorter than $\mathbf p_{t}^{j}$ is
\begin{equation}
    P({\Delta {\hat w}_j}\geq0)=\Phi\left(\frac{\Delta w_j}{\sqrt{\sigma_{\mathbf {p}_t^*}^2+\sigma_{\mathbf p_{t}^{j}}^2}}\right),
\end{equation}
where $\Phi(\cdot)$ is the cumulative distribution function of the standard Gaussian distribution.

The accuracy of path planning, denoted as \( {\mathbf Q}(\mathbf p_t^*=\mathbf {\hat p}_t^*) \), can be formalized as the probability of $\mathbf p_t^*=\mathbf {\hat p}_t^*$. Under the simplifying assumption that path comparisons are independent, it can be expressed as the joint probability that $\mathbf{p}_t^*$ is better than all the alternative paths $\mathbf{p}_{t}^{j} \ne \mathbf{p}_t^*$, as follows:
\begin{align}
        \label{Q}
        {\mathbf Q}(\mathbf p_t^*=\mathbf {\hat p}_t^*)=\prod\limits_{{\rm{\mathbf p_{t}^{j}}} \ne \mathbf {p}_t^*}\Phi\left(\frac{\Delta w_j}{\sqrt{\sigma_{\mathbf {p}_t^*}^2+\sigma_{\mathbf p_{t}^{j}}^2}}\right).
\end{align}
It can be seen from \eqref{Q} that a smaller value of ${\sigma_{\mathbf{p}_t^*}^2 + \sigma_{\mathbf{p}_{t}^{j}}^2}$ leads to a higher value of $\mathbf{Q}(\mathbf{p}_t^* = \hat{\mathbf{p}}_t^*)$. 
In practice, candidate paths are not completely independent, as they often overlap in vertices and edges and may share correlated perturbations. Modeling these correlations would require evaluating a full joint distribution, which is generally intractable in closed form. Therefore, to keep the analysis manageable, we assume independence as an approximation, which enables us to derive a closed-form expression showing how reducing the uncertainty on the paths most competitive with the optimum increases the likelihood of recovering the correct route.

In our proposed SemCom framework, the transmission accuracy of a UAV-covered area can be improved by increasing $\delta_k$, which reduces the weight estimation error of certain vertices and thereby decreases $\sigma_{\mathbf{p}_{t}^{j}}^2$. It is therefore important to analyze which candidate paths, through the reduction of their associated variance $\sigma_{\mathbf{p}_{t}^{j}}^2$, contribute most significantly to improving $\mathbf{Q}(\mathbf{p}_t^* = \hat{\mathbf{p}}_t^*)$.

The contribution of reducing $\sigma_{\mathbf{p}_{t}^{j}}^2$ depends on the corresponding weight gap $\Delta w_j$ between $\mathbf{p}_t^*$ and $\mathbf{p}_{t}^{j}$. To characterize this relationship, we analyze the rate of change of $\mathbf{Q}$ with respect to $\sigma_{\mathbf{p}_{t}^{j}}^2$. Specifically, to analyze the effect of variance reduction on a particular candidate path, we approximate the global change in~(\ref{Q}) by examining the change in a single term. This approximation assumes that the influence of each term is averaged in the overall probability, allowing us to estimate the influence of variance reduction on~(\ref{Q}) in a tractable way.

Proposition 1 identifies the condition under which the rate of change of $\mathbf{Q}$ is maximized, thereby revealing which candidate paths are more sensitive and thus more valuable for targeted variance reduction.
\begin{proposition}
In the Gaussian case, the impact of reducing $\sigma^2_{\mathbf{p}_{t}^{j}}$ on $\mathbf{Q}(\mathbf{p}_t^* = \hat{\mathbf{p}}_t^*)$ is maximized when
\begin{align}
    {\sigma^2_{\mathbf{p}_{t}^{j}}} = \Delta w_j^2-\sigma^2_{\mathbf{p}_t^*}.
\end{align}
\end{proposition}
\renewcommand\qedsymbol{$\blacksquare$}
\begin{proof}
Consider the $j$-th term of the product in~(\ref{Q}):
\begin{align}
    f(\sigma_{\mathbf{p}_{t}^{j}}^2) = \Phi\left( \frac{\Delta w_j}{\sqrt{\sigma^2_{\mathbf{p}_t^*} + \sigma^2_{\mathbf{p}_{t}^{j}}}} \right). 
\end{align}
Taking its partial derivative with respect to $\sigma^2_{\mathbf{p}_{t}^{j}}$ gives
\begin{align}
    \left| \frac{\partial f}{\partial \sigma^2_{\mathbf{p}_{t}^{j}}} \right|
    = \frac{1}{2} \cdot \frac{|\Delta w_j|}{(\sigma^2_{\mathbf{p}_t^*} + \sigma^2_{\mathbf{p}_{t}^{j}})^{3/2}} \cdot \phi\left( \frac{\Delta w_j}{\sqrt{\sigma^2_{\mathbf{p}_t^*} + \sigma^2_{\mathbf{p}_{t}^{j}}}} \right), 
\end{align}
where $\phi(\cdot)$ is the probability density function of the standard normal distribution.

Let $z = \frac{\Delta w_j}{\sqrt{\sigma^2_{\mathbf{p}_t^*} + \sigma^2_{\mathbf{p}_{t}^{j}}}}$. 
Then (12) becomes
\begin{align}
    \left| \frac{\partial f}{\partial \sigma^2_{\mathbf{p}_{t}^{j}}} \right|
    = \frac{1}{2} \cdot \frac{1}{\sqrt{\sigma^2_{\mathbf{p}_t^*} + \sigma^2_{\mathbf{p}_{t}^{j}}}} \cdot |z| \cdot \phi(z). 
\end{align}
Define the function
\begin{align}
f(z) = |z| \cdot \phi(z) = \frac{|z|}{\sqrt{2\pi}} e^{-z^2 / 2}.
\end{align}
Since $f(z)$ reaches its maximum at $|z| = 1$, the sensitivity is maximized when
\begin{align}
    \left| \frac{\Delta w_j}{\sqrt{\sigma^2_{\mathbf{p}_t^*} + \sigma^2_{\mathbf{p}_{t}^{j}}}} \right| = 1, 
\end{align}
which implies
$\sigma^2_{\mathbf{p}_{t}^{j}} = \Delta w_j^2-\sigma^2_{\mathbf{p}_t^*}. 
$
\end{proof}

Proposition 1 identifies which candidate paths are more effective in improving $\mathbf{Q}(\mathbf{p}_t^* = \hat{\mathbf{p}}_t^*)$ when their associated variance is reduced. 
From the proof of Proposition 1, we extract $f(z) = |z| \cdot \phi(z)$,
which characterizes how rapidly $\mathbf{Q}(\mathbf{p}_t^* = \hat{\mathbf{p}}_t^*)$ changes with respect to the variance $\sigma^2_{\mathbf{p}_t^j}$. A larger $f(z)$ implies that reducing the variance of path $\mathbf{p}_{t}^{j}$ leads to a greater improvement in $\mathbf{Q}(\mathbf{p}_t^* = \hat{\mathbf{p}}_t^*)$. Therefore, $f(z)$ can be used as a heuristic path-wise priority metric for guiding the allocation of transmission resources. The derivation uses the independence assumption in (10). Even without spatial correlations, the metric is still useful because it highlights the most sensitive paths. We denote $f(z)$ by $\xi$ for brevity in the remainder of this section.

\subsection{Problem Solution}
Consider that the UGVs typically have less stringent power and compute constraints. Therefore, based on Proposition 1, we design a transmission strategy by combining the Monte Carlo method and Local Betweenness Centrality (LBC) to guide the improvement of ${\mathbf Q}(\mathbf p_t^*=\mathbf {\hat p}_t^*)$.

In the proposed method we first compute the LBC. 
Betweenness centrality measures the frequency with which a given node appears in all shortest paths within a graph to evaluate node importance. Computing betweenness centrality requires determining the shortest paths for all node pairs and counting how often each node is traversed, leading to a significant computational burden. 
Therefore, we generate a set of paths between the predefined source and destination points using a Monte Carlo method and then identify all paths for which \(\xi \ge h\).
Subsequently, we count the frequency of being traversed for each node in these paths and employ this metric in place of the original betweenness centrality for evaluating node importance, which we refer to as LBC. The LBC of $u$ is denoted by $\psi_{u,h}$. 
In the Monte Carlo method, paths are generated by selecting edges according to their weights at each node. Under the constraint that each vertex is visited at most once per path, the probability of transitioning from any vertex $u$ to an adjacent vertex $v$ is expressed as
\begin{equation}
\begin{aligned}
\label{path}
p(u,v)=\frac{1/(w(u,v)+\iota )}{\sum\limits_{v'\in \mathcal V(u)}{1/(w(u,v')+\iota )}},
\end{aligned}
\end{equation}
where $\mathcal V(u)$ is the set of vertices adjacent to $u$ and $\iota$ is a very small coefficient to prevent division by 0.

Then we compute the sum of the LBC of all vertices in the observation area of the \( k \) UAV $\Psi_k$, denoted as $\Psi_k$ and
\begin{equation}
\begin{aligned}
\label{psi}
\Psi_k=\sum\limits_{u \in {G_{t,k}}} {{\psi_{u,h}}},
\end{aligned}
\end{equation}
where $G_{t,k}$ is the set of points in the map observed by the $k$th UAV.
For any given path \( \mathbf{p_{t}^{j}} \), the expected number of its vertices within the \( G_{t,k} \) region is \(e_{k,\mathbf{p_{t}^{j}}}=\mathbb E(|G_{t,k} \cap {\mathbf{p_{t}^{j}}}|) \). 
The connection among $\Psi_k$, $\xi$, and \(e_{k,\mathbf{p_{t}^{j}}}\) can be described by the following mixture Gaussian model:
\begin{equation}
\begin{aligned}
&P(\psi_k,e_{k,\mathbf{p_{t}^{j}}} |\xi)\\
=&w(\xi)P_\mathrm{c}(\psi_k,e_{k,\mathbf{p_{t}^{j}}} )+(1-w(\xi))P_\mathrm{d}(\psi_k,e_{k,\mathbf{p_{t}^{j}}} ).
\end{aligned}
\end{equation}
where $P_\mathrm{c}(\psi_k,\mathbb E(|G_{t,k} \cap {\mathbf{p_{t}^{j}}}|)$ is a positively correlated Gaussian distribution, $P_\mathrm{d}(\psi_k,\mathbb E(|G_{t,k} \cap {\mathbf{p_{t}^{j}}}|)$ is the independent distribution, and $w(\xi)$ is a weight controlling the mixing ratio of the two distributions.
In particular, as $\xi$ increases, the correlation between $\Psi_k$ and $\mathbb E(|G_{t,k} \cap {\mathbf{p_{t}^{j}}}|)$ tends to be positive, whereas as $\xi$ decreases, $\Psi_k$ and $\mathbb E(|G_{t,k} \cap {\mathbf{p_{t}^{j}}}|)$ tend to become independent.
Then, we set $\delta_k$ based on \( \psi_k \). Specifically, in regions with higher \( \psi_k \), we set a larger $\delta_k$ to ensure that more vertices in paths with higher \( \xi \) can be transmitted more accurately.
Therefore, for the $k$-th UAV, $\delta_k$ is computed by 
\begin{equation}
\begin{aligned}
\label{d}
{{\delta }_{k}}=\frac{1}{1+e^{\varrho\cdot{(\Psi_h-\Psi_k)}}},
\end{aligned}
\end{equation}
where $\Psi_h$ is the sum of the betweenness centrality of all points along the paths that \( \xi>h \) and $\varrho$ is the normalized coefficient. 

In this way, the proposed method ensures that regions containing more vertices in paths with higher $\xi$ are transmitted more accurately, thereby reducing the error of higher $\xi$ paths to a greater extent. According to Proposition 1, this method enhances ${\mathbf Q}(\mathbf p_t^*=\mathbf {\hat p}_t^*)$. In particular, all Monte Carlo sampling and aggregation run on the UGV, whereas the UAV-side cost is governed by the sparsification ratio $\delta_k$ in (19). The UGV typically has more energy and computing resources than lightweight UAVs, so concentrating computation on the UGV is a reasonable deployment choice.

To further reduce the online computation cost while preserving the semantics-driven allocation strategy, we introduce Partial offline precomputation and Windowed online update to make the Algorithm lightweight.
We precompute an LBC map $\psi^{\mathrm{off}}_{u,h}$ on the UGV based on a prior map snapshot. This is done by running the Monte Carlo sampling defined in~(16) over the full map once, and storing the resulting node importance values. This offline result will be reused in subsequent decision epochs.

At each decision epoch $t$, we compute the current shortest path $\hat p_t^\ast$ on $\hat G_t$. The update window is defined as
\begin{equation}
\begin{aligned}
W_t = \mathcal N_r(C_t \cup F_t),
\end{aligned}
\end{equation}
where $C_t$ denotes the path corridor, i.e., all vertices within a fixed radius $r_c$ of the current shortest path $\hat p_t^\ast$, $F_t = \{u \mid \hat G_t(u) \neq \hat G_{t-1}(u)\}$ is the frontier of updated cells, and $\mathcal N_r(S)$ denotes the set of vertices within graph distance $r$ of $S$. In particular, the Monte Carlo path sampling in~(16) is performed only inside $W_t$.
The resulting online increment $\Delta\psi^{\mathrm{on}}_{u,h}(t)$ is added to the offline $\psi^{\mathrm{off}}_{u,h}$ to obtain the working LBC:
\begin{equation}
\begin{aligned}
\tilde\psi_{u,h}(t) = \psi^{\mathrm{off}}_{u,h} + \Delta\psi^{\mathrm{on}}_{u,h}(t).
\end{aligned}
\tag{22}
\end{equation}
The $\tilde\Psi_k(t)$ is then computed by~(17) with $\tilde\psi_{u,h}(t)$, and the coding ratio $\delta_k$ is still determined by~(19) after replacing $\Psi_k$ with $\tilde\Psi_k(t)$.  
This MC-LBC-RT realization keeps the decision logic unchanged, but makes the runtime cost proportional to $|W_t|$ instead of the full map size~$|V|$. The processing of the proposed method is described in Algorithm 1.

\begin{algorithm}[t]
    \caption{Multi-transceiver coding strategy with partial offline precomputation and windowed Monte Carlo sampling}
    \begin{algorithmic}[1]
    \State \textbf{Input:} $\mathbf{\hat G}_t$, $\mathbf{\hat G}_{t-1}$, threshold $h$, radius parameters $r_c, r$
    \State \textbf{Output:} $\{\delta_k\}$
    \State \textbf{Offline:} On $\mathbf{\hat G}_{t-1}$, run Monte Carlo sampling by~(\ref{path}) over the full map to compute and store $\psi^{\mathrm{off}}_{u,h}$.
    \State \textbf{Step 1: Generate Paths using Monte Carlo Method in update window}
        \State Compute current shortest path $\hat p_t^\ast$ on $\mathbf{\hat G}_t$.
        \State Let $C_t$ be the path corridor (vertices within radius $r_c$ of $\hat p_t^\ast$).
        \State Let $F_t = \{u \mid \hat G_t(u) \neq \hat G_{t-1}(u)\}$ be the frontier of updated cells.
        \State Define $W_t = \mathcal N_r(C_t \cup F_t)$, where $\mathcal N_r(S)$ denotes the set of vertices within graph distance $r$ of $S$ (in grid maps, within $r$ cells).
    \For{$j = 1$ to Traverse time}
        \State Generate path $\mathbf p_{t}^{j}$ \emph{restricted to $W_t$} by~(\ref{path})
        \For{each UAV $k$ in $n$}
            \State Compute local increment $\Delta\psi^{\mathrm{on}}_{u,h}(t)$ for visited $u \in W_t$ if $\xi \ge h$
            \State Compute $\tilde\Psi_k$ by~(\ref{psi}) using $\tilde\psi_{u,h}(t) = \psi^{\mathrm{off}}_{u,h} + \Delta\psi^{\mathrm{on}}_{u,h}(t)$
        \EndFor
    \EndFor
    \State \textbf{Step 2: Compute $\delta_k$}
    \For{$k = 1$ to $n$}
        \State Compute $\delta_k$ by~(\ref{d}) with $\Psi_k$ replaced by $\tilde\Psi_k$
    \State \textbf{Return} $\tilde\Psi_k$ and $\{\delta_k\}$
    \EndFor
\end{algorithmic}
\end{algorithm}

\section{Adaptive SemCom Transceiver for Path Planning}
\begin{figure*}[htp]
	\centering
        \setlength{\abovecaptionskip}{-0pt}
        \setlength{\belowcaptionskip}{-0pt}
	\includegraphics[width=0.9\textwidth]{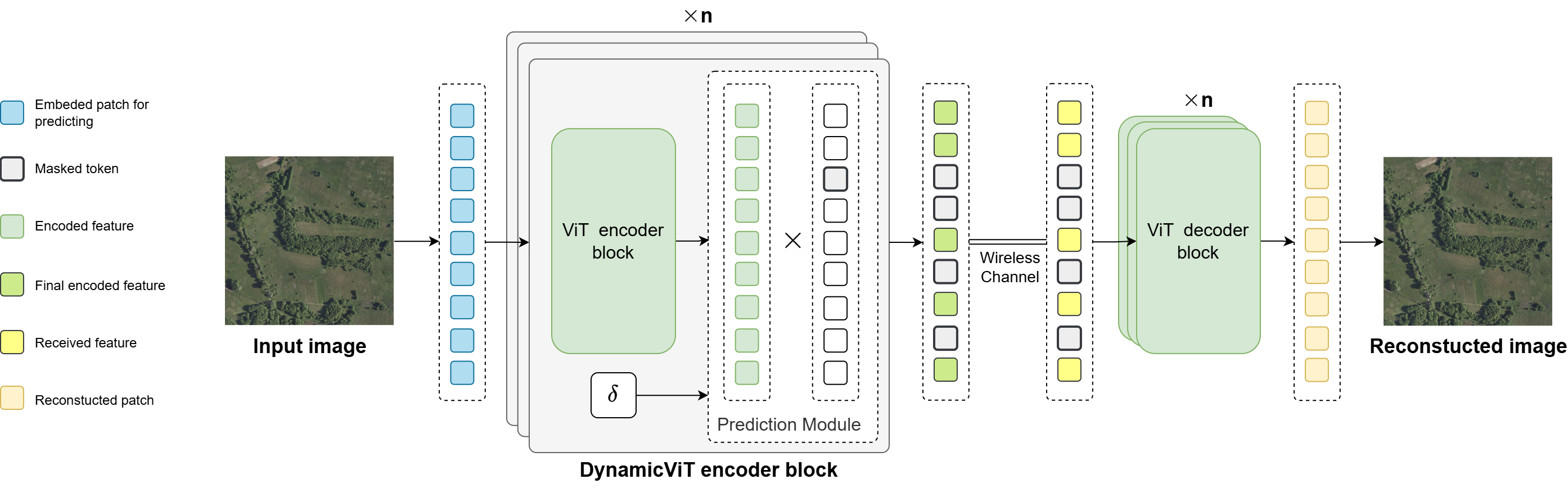}
  	\caption{Dynamic Sparsification ViT Based SemCom encoder and decoder network.}
	\label{img2}
\end{figure*}
Building upon the transmission strategy proposed in Section IV, we design a SemCom coding transceiver named Path-SC to dynamically adjust the image transmission strategies. The Path-SC transceiver employs a training framework similar to that of the Masked Autoencoder (MAE) \cite{MAE} to recover the semantic information from adjacent patches. Meanwhile, it integrates a DynamicViT block \cite{dn} to evaluate the importance of each patch and adapt the transmission strategy based on the parameter \(\delta_k\).

\subsection{Structure of the Semantic Encoder and Decoder}
As shown in Fig. 2, the Path-SC semantic encoder comprises \( n \) DynamicViT modules, each retaining the full structure of a traditional ViT while incorporating an additional prediction module at the end. Specifically, the input image \( \mathbf{x}_{t}^{k} \in \mathbb{R}^{h \times d \times c} \) is first partitioned into \( L \) patches of size \( b \times b \). Each patch \( w_{tl} \) (\( l \in [1, N] \)) is then flattened into a one-dimensional vector and linearly projected to form the token \( \mathbf{t}_{tl} \in \mathbb{R}^{L_s \times 1} \), where \( L_s \) denotes the token length.  

These tokens are sequentially processed through a transformer block, enabling global attention computation and feature extraction. The block's output is fed into a lightweight multilayer perceptron-based prediction module, which evaluates token importance relative to the original image. Based on this evaluation and a dynamic adjustable parameter \( \delta \), the module dynamically selects high-importance tokens for further processing in the next ViT block, while low-importance tokens compute self-attention only. This design reduces the overall computational complexity.

The Path-SC semantic decoder comprises \( n \) ViT decoder blocks, each embedding a transformer module. These blocks process all tokens to reconstruct the input image by predicting pixel values per token. Contextual information from neighboring regions mitigates noise, enhancing image semantics. The final decoder layer employs a linear projection to reshape vectors into \( b \times b \) patches, reconstructing \( \mathbf {\hat x}_{t}^{k} \). The proposed semantic decoder optimizes the average mean squared error between the original image \( \mathbf{x}_{t}^{k} \) and the reconstructed image \( \mathbf{\hat{x}}_{t}^{k} \) as its loss function, given as follows:
\begin{equation}
\mathcal {L}_{MSE} = \frac {1}{N}\sum _{i=1}^{N} \|{\mathbf x_{t}^{k}-  { \mathbf {\hat x}_{t}^{k}}}\|^2.
\end{equation}

\subsection{Training Process}
To improve the accuracy of reconstructed images in SemCom, we employ a masking technique during model training \cite{MAE}. Specifically, only a subset of image patches is fed into the encoder, while the remaining patches are masked. The decoder then reconstructs the full image from the visible patches.
During training, the prediction module is optimized alongside the transceiver. To mitigate the impact of sparsification, we adopt knowledge distillation as that in \cite{DeiT}, where a fully trained semantic encoding network (teacher model) guides the proposed transceiver training using the same loss function:
\begin{align}
\label{loss}
 \mathcal L ={}& {{\mathcal L}_{MSE}} + {{\lambda}_\mathbf {KL}}\mathbf {KL}(\mathbf {\hat 
 x}_{t}^{k}|| {\mathbf{\hat x}_{t}^{\prime k}}) + {\lambda }_\mathbf {\delta_k}{{\mathcal L}_{{\delta_k}}} \nonumber\\
&+ {{\lambda }_{distill}}\frac{\sum_{t=1}^{T}{\sum_{l=1}^{N}{\mathbf {\hat D}_{l}^{t,M}}}{{({{\mathbf t}_{tl}}-\mathbf t'_{tl})}^{2}}}{\sum_{t=1}^{T}{\sum_{l=1}^{N}{\hat{\mathbf D}_{l}^{t,M}}}},
\end{align}
where $\mathbf {KL}({{\mathbf {\hat x}}_{t}^{k}}|| {{\mathbf {\hat x}}_{t}^{\prime k}})$ is the KL divergence between \(\mathbf{x}_{t}^{k}\) and the teacher model output \(\mathbf {\hat x}_{t}^{\prime k}\). \(\mathbf{\hat{D}_{l}^{t,M}}\) is a binary variable indicating whether the input sample at time \(t\) retains the \(k\)-th token at the \(M\)-th sparsification stage. \(\lambda_\mathbf{KL}\), \(\lambda_{distill}\), and \(\lambda_\mathbf{\delta_k}\) are weighting coefficients. In particular, the loss function \(\mathcal{L}_{\delta_k}\) in (\ref{loss}) is defined as
\begin{equation}
\begin{aligned}
{{\mathcal L}_{{\delta_k}}}=\frac{1}{TM}\sum\nolimits_{t=1}^{T}{\sum\nolimits_{m=1}^{M}}\left({\delta_k}-\frac{1}{N}\sum\nolimits_{l=1}^{N}{\mathbf {\hat D}_{l}^{t,m}} \right).
\end{aligned}
\end{equation}
The training process is shown in Algorithm 2. 
\begin{algorithm}[t]
    \caption{Training Process}
    \label{alg2}
    \begin{algorithmic}[1]
        \State Initialization: Training data $\mathcal W$, learning rate $\gamma$.     
        \Repeat
        \State \parbox[t]{\dimexpr\linewidth-\algorithmicindent}{Input: Training data $\mathbf{x}_{t}^{k}$ and learning rate $\Lambda$.}
        \State \parbox[t]{\dimexpr\linewidth-\algorithmicindent}{Calculate the proportion of the retained token $\delta_k$ based on Algorithm 2.}
        \State \parbox[t]{\dimexpr\linewidth-\algorithmicindent}{Embed $\mathbf x_{t}^{k} \to \{x_{t}^{k1},x_{t}^{k2},\cdots,x_{t}^{km}\}$.}
        \State \parbox[t]{\dimexpr\linewidth-\algorithmicindent}{$\mathbf x_{t}^{k} \to  {\mathcal{ E}_e}\left ({{\mathbf {x}_{t}^{k}}; { \boldsymbol \delta_k, \alpha}} { }\right) \to \mathbf s_{t}^{k} \to \mathbf {\hat s}_{t}^{k} \to {\mathcal{ E}_d}\left ({{\mathbf {\hat s}_{t}^{k}}; { \boldsymbol \delta_k, \alpha}} { }\right) \to \mathbf {\hat x}_{t}^{k}$.}
        \State \parbox[t]{\dimexpr\linewidth-\algorithmicindent}{Calculate the loss $\mathcal {L}_{MSE} = \frac {1}{N}\sum _{i=1}^{N} d\left ({\mathbf x_{t}^{k}, {\mathbf {\hat x}}_{t}^{k}}\right).$}
        \State \parbox[t]{\dimexpr\linewidth-\algorithmicindent}{ Calculate the distillation loss function based on (20).}
        \State \parbox[t]{\dimexpr\linewidth-\algorithmicindent}{Update $\alpha$ by $\mathcal L$.}
    \Until{$t = T$}
    \end{algorithmic}
\end{algorithm}

\section{Simulation Results and Discussion}
In this section, we conduct simulations to evaluate the performance of the proposed Path-SC in terms of path planning accuracy, end-to-end transmission time, and demand transmission rate.

\subsection{Simulation Setup}
\begin{figure*}[htp]
	\centering
        \setlength{\abovecaptionskip}{-0pt}
        \setlength{\belowcaptionskip}{-0pt}
        \vspace{-0cm}
	\includegraphics[width=0.95\textwidth]{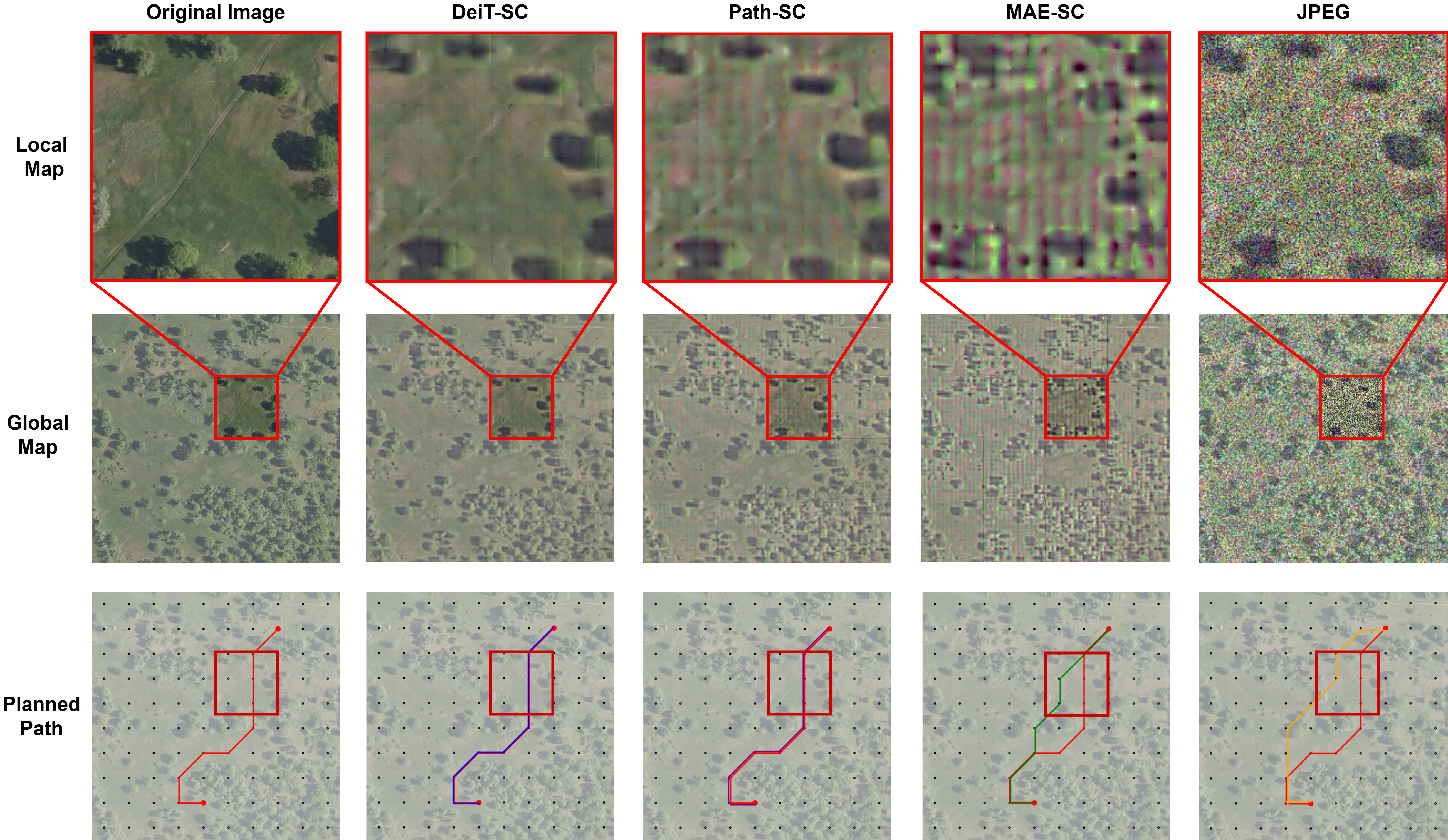}
  	\caption{Comparison of Path-SC, DeiT-SC and MAE-SC reconstructed images and path planning result under SNR=12 dB.}
	\label{img2}
\end{figure*}
We employ the LandCover.ai dataset \cite{lancover} to train the transceiver. 
The dataset consists of 41 high-resolution aerial images of rural Poland, depicting buildings, forests, water bodies, and roads. 
Specifically, 33 orthophotos are provided with a size of approximately $9000 \times 9500$ pixels, and 8 orthophotos with a size of approximately $4200 \times 4700$ pixels, covering a total area of 216\,km$^2$.
Its varied and realistic geographic features make it well-suited for representing typical UAV/UGV path planning environments.
We consider a single UGV and 16 UAVs. The UGV executes the path planning, while UAVs sense and transmit images via the proposed Path-SC and three benchmarks. All model training is performed by fine-tuning a pre-trained DeiT model\cite{DeiT}. The encoder transformer comprises 12 layers with 12 attention heads, while the decoder transformer consists of 8 layers with 12 attention heads. In the training process, we employ the Adam optimizer\cite{adam} with a learning rate of ${10}^{-4}$, the parameter $\delta$ is fixed at 0.7, and the batch size is 16. The program code is implemented in PyTorch and executed on an RTX A6000 GPU.
In our simulations, we adopt the full 12-layer ViT encoder to reflect the capability of relatively strong UAV platforms, while in practical deployment the encoder depth and the sparsity ratio \(\delta\) can be adjusted according to the computational characteristics of the specific UAV.

In the simulation, we first segment the original ultra-high-resolution image into multiple smaller images, each with a size of 1500×1500 pixels. The smaller images serve as task ground maps for the path-planning process. Each task ground map is further divided into 16 subregions, each corresponding to the observation range of a UAV. 
Subsequently, we transmit these smaller images using three different transceivers. The received images are then used to construct three distinct traversable graphs based on the algorithm proposed in \cite{trav}. According to \cite{trav}, the edges identified as impassable in the original image are marked as labels for path planning.
Based on these traversable graphs, we employ Dijkstra's algorithm to compute the shortest paths, which serves as the standard shortest path algorithm to isolate the impact of semantic transfer.
Finally, the performance of the Path-SC and three benchmarks (two semantic coding benchmarks and JPEG) in path planning is evaluated by comparing the shortest paths generated from the images transmitted by each transceiver with the reference path derived from the original images.

The two semantic coding benchmark models that we consider are: 
\begin{itemize}
\item
\textit{DeiT-SC:} 
A SemCom model fine-tuned from the original Deit\cite{DeiT} without the dynamic sparsity module and the MASK-based training method.
\item
\textit{MAE-SC:} 
A SemCom model based on the original MAE \cite{MAE} that applies fixed-ratio masking of patches without employing the dynamic sparsification method. 
\end{itemize}
For a fair comparison, all parameters of DeiT-SC and MAE-SC, except for those in the prediction module, are identical to those of Path-SC.

\subsection{Path Planning Performance Comparisons}
We compare the images received by Path-SC and two benchmarks, along with the planned paths on these images. We then evaluate the performance of Path-SC by analyzing the differences in the total path weight error between the planned trajectories on the received images and the original reference path, and the frequency of times the planned path traverses non-traversable regions.

Fig. 3 presents a comparison of the performance between the Path-SC and two benchmarks in transmitting images and planning paths under an SNR=12 dB. We select a forest image as the sample for demonstration.  
In particular, the \emph{Local Map} corresponds to the image transmitted by a single UAV, while the \emph{Global Map} is formed by fusing images collected by all UAVs.
It can be observed that DeiT-SC effectively preserves the image features, ensuring that the received image closely resembles the original and enabling the planned path to align precisely with the reference path. In contrast, Path-SC introduces distortions due to model sparsification. However, it retains essential semantic information at critical locations, ensuring that the planned path remains consistent with the reference path. MAE-SC transmits only partial features, leading to significant distortions in the reconstructed image and causing the planned path to deviate from the reference path. In contrast, the JPEG method introduces severe noise artifacts under the given channel conditions, which confuse critical scene details. Consequently, the planned path exhibits the largest divergence from the reference path among all methods.
This comparison validates that Path-SC effectively preserves the key semantic information required for path planning while reducing transmission and computing consumption.
\begin{table*}[t]
    \setlength{\abovecaptionskip}{0pt}
    \setlength{\belowcaptionskip}{0pt}
    \small
    \label{tab:runtime_comparison}
    \renewcommand{\arraystretch}{1.5}
    \centering
    \caption{Comparison of the proposed Path-SC with other benchmarks in terms of data volume and runtime decomposition.}
    \begin{tabular}{|c|c|c|c|c|}
        \hline
        \textbf{Method} & \textbf{Data Volume (kb)} & \textbf{UGV Computation Time (ms)} & \textbf{End to End Transmission Time (ms)} & \textbf{Total Time (ms)} \\
        \hline
        MAE-SC   &  \,\,296.4   & 32.75 &  45.19     &  77.94   \\
        DeiT-SC  &  \,\,988.3   & 32.75 &  50.91     &  83.66   \\
        Path-SC  &  \,\,563.2   & 58.92 &  48.38     &  107.30   \\
        JPEG     &  \,\,1216.3  & 32.75 &  52.45     &  85.20   \\
        \hline
    \end{tabular}
\end{table*}
\begin{figure}[htp]
	\centering
        \setlength{\abovecaptionskip}{-0pt}
        \setlength{\belowcaptionskip}{-0pt}
        \vspace{-0cm}
	\includegraphics[width=0.45\textwidth]{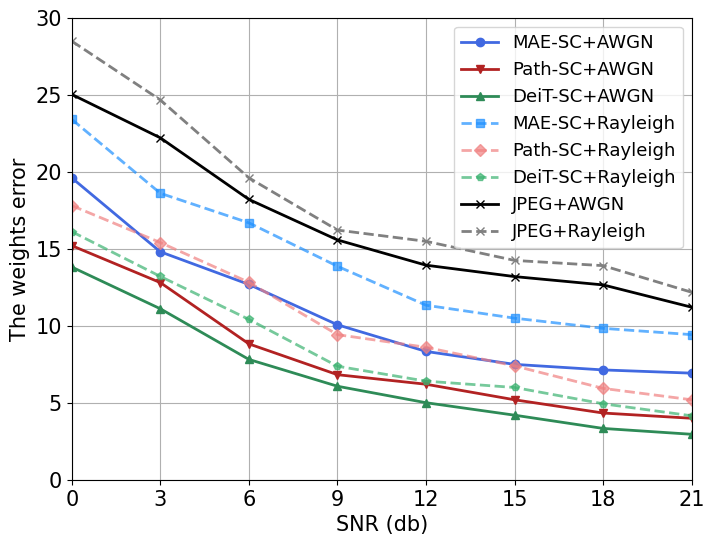}
  	\caption{Total weight error of Path-SC and two other benchmarks under AWGN and Rayleigh fading channels under different SNR.}
	\label{img2}
\end{figure}
\begin{figure}[htp]
	\centering
        \setlength{\abovecaptionskip}{-0pt}
        \setlength{\belowcaptionskip}{-20pt}
        \vspace{-0cm}
	\includegraphics[width=0.45\textwidth]{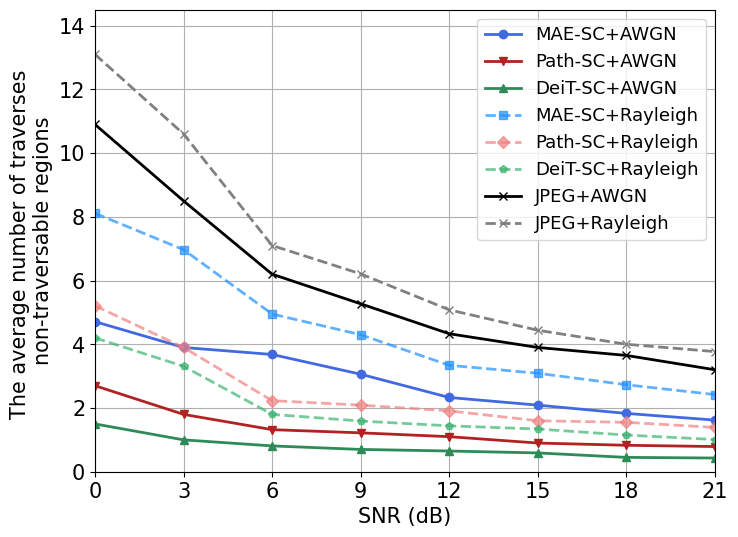}
  	\caption{Number of times the planned path traverses non-traversable region of Path-SC and two other benchmarks under AWGN and Rayleigh fading channels under different SNR.}
	\label{img2}
\end{figure}

We then evaluate the performance of the Path-SC and two benchmarks in terms of the total path weight error and the number of times the planned path traverses non-traversable regions. 
Fig. 4 shows the total weight error of Path-SC and two benchmarks under both AWGN and Rayleigh fading channels across different SNR values. As the SNR increases, the total weight error for all methods decreases. DeiT-SC consistently achieves the lowest error, while JPEG produces the highest error across both channel types. The performance gap between AWGN and Rayleigh fading channels is evident, especially at lower SNRs.
This result show that the Path-SC, which applies model sparsification while maintaining model performance, shows a clear advantage over MAE-SC, particularly at higher SNRs. MAE-SC experiences significant degradation under Rayleigh fading, further amplifying its performance limitations. This highlights the effectiveness of Path-SC in maintaining a balance between sparsity and performance, while the performance JPEG illustrates the limitations of traditional coding schemes in noisy channels.

Similarly, Fig. 5 shows the number of times the planned path traverses non-traversable regions for Path-SC and three benchmarks under different SNR values. The overall trend resembles that of Fig. 4, with the number of traverses decreasing as SNR increases. Notably, the variations in DeiT-SC and Path-SC remain relatively moderate compared to the significant fluctuations observed in MAE-SC. This can be attributed to the ability of SemCom to retain critical semantic features of the environment, such as obstacle contours and sizes, thereby mitigating the effect of noise on path planning. In contrast, MAE-SC loses the essential semantic information during transmission, making path planning more vulnerable to noise-induced errors. Compared with the SemCom methods, JPEG produces the highest number of non-traversable regions traversed across all SNR values.

Fig. 6 shows the weight error of the four schemes on the full dataset and on the five urban images under AWGN and Rayleigh channels. DeiT‑SC gives the lowest error, Path‑SC is the next best, MAE‑SC is worse, and JPEG is the worst across all settings. On the urban subset, the error increases for every method, but the relative ordering and overall trend remain the same as in the global case. The SemCom methods have a moderate increase, while JPEG shows a clear jump, indicating that semantic encoders keep task‑relevant structure better than pixel‑level coding. These results show that our framework maintains accuracy in urban environments and follows the same performance trend observed on the full dataset.

\begin{figure}[htbp]
    \centering
        \setlength{\abovecaptionskip}{-0pt}
        \setlength{\belowcaptionskip}{-0pt}
        \vspace{-0cm}
    \includegraphics[width=0.48\textwidth]{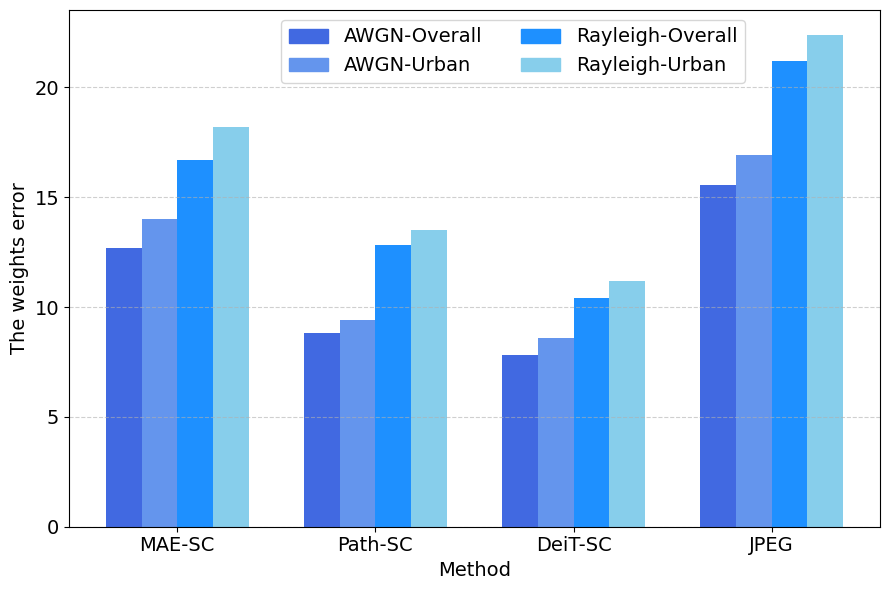}
    \caption{Weight error comparison on the full dataset and the urban subset under AWGN and Rayleigh channels.}
    \label{img2}
\end{figure}

Table I presents the average data volume and runtime decomposition of the four methods under a bandwidth of 10 MHz, evaluated over 100 trial images. In terms of data volume, JPEG produces the largest payload, while three SemCom methods significantly reduce the data size. The UGV computation time refers to the processing on the UGV side, which mainly includes path planning and, for Path-SC, the extra adaptive computing for the communication strategy. For the benchmark methods, the UGV computation time stays nearly constant at a low level, while Path-SC shows a noticeably higher value because of the added operations. For transmission time, SemCom methods send much less data, so even with the encoder and decoder included, they are still faster, while JPEG is the slowest because of its large data volume. Overall, considering the total time, MAE-SC has the lowest latency, JPEG and DeiT-SC show comparable levels, and Path-SC is moderately higher because of the extra computation. Nevertheless, the increase in total time for Path-SC is not significant compared with the other methods, and in UAV/UGV cooperative tasks such additional latency is minor and can be tolerated, especially given its advantages in reducing computation cost and improving robustness in path planning.

\section{Conclusion}
In this paper, we propose a SemCom framework for UAV/UGV cooperative path planning, which reduces data transmission by focusing on task-relevant semantic information while preserving path planning accuracy. We theoretically proved the conditions for identifying critical regions that are essential for accurate path planning and designed a transmission strategy that prioritizes these regions. Guided by this condition, we developed a SemCom transceiver tailored to the dynamic and resource-constrained UAV/UGV environment. Simulation results demonstrate that the proposed framework significantly reduces computing overhead on UAVs and maintains high path planning accuracy, thereby contributing to overall mission success.
This study provides insights for the future integration of SemCom and control systems, encouraging further research on dynamic encoding strategies and transmission mechanisms to tackle the challenges in autonomous vehicle operations. Future work could extend this framework to handle large datasets and multimodal inputs, which would strengthen the scalability and applicability of the proposed framework in complex cooperative tasks.


%


\ifCLASSOPTIONcaptionsoff
  \newpage
\fi



%

\bibliographystyle{IEEEtran}
\bibliography{IEEEexample}
\vfill
\begin{IEEEbiography}
[{\includegraphics[width=1in,height=1.25in,clip,keepaspectratio]{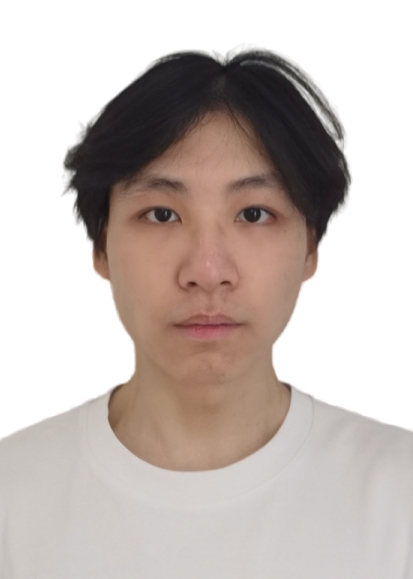}}]
{Fangzhou Zhao} (Graduate Student Member, IEEE) received the B.Eng. degree from the School of Information Engineering, Zhejiang University of Technology, Hangzhou, China, in 2018, and the M.Sc. degree from the James Watt School of Engineering, University of
Glasgow, Glasgow, U.K.,in 2020. He is currently working toward the Ph.D. degree with James Watt School of Engineering, University of Glasgow, Glasgow, U.K. His research interests include deep learning in wireless communication and semantic communication.
\end{IEEEbiography}

\begin{IEEEbiography}
[{\includegraphics[width=1in,height=1.25in,clip,keepaspectratio]{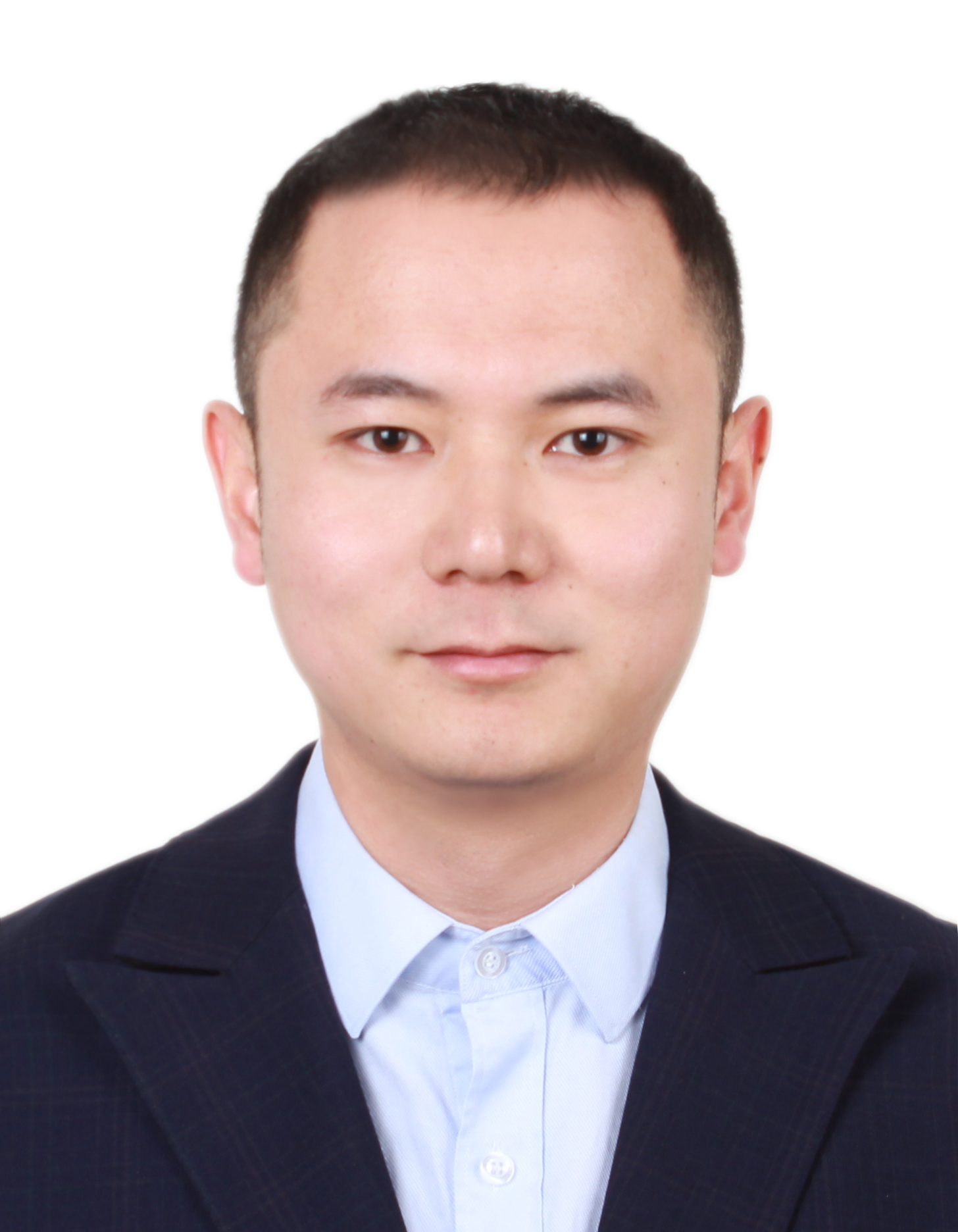}}]
{Yao Sun} (Senior Member, IEEE) is currently a Lecturer with James Watt School of Engineering, the University of Glasgow, Glasgow, UK. Dr Sun has won the IEEE Communication Society of TAOS Best Paper Award in 2019 ICC, IEEE IoT Journal Best Paper Award 2022 and Best Paper Award in 22nd ICCT. His research interests include intelligent wireless networking, semantic communications, blockchain system, and resource management in next generation mobile networks. Dr. Sun is a senior member of IEEE.
\end{IEEEbiography}

\begin{IEEEbiography}
[{\includegraphics[width=1in,height=1.25in,clip,keepaspectratio]{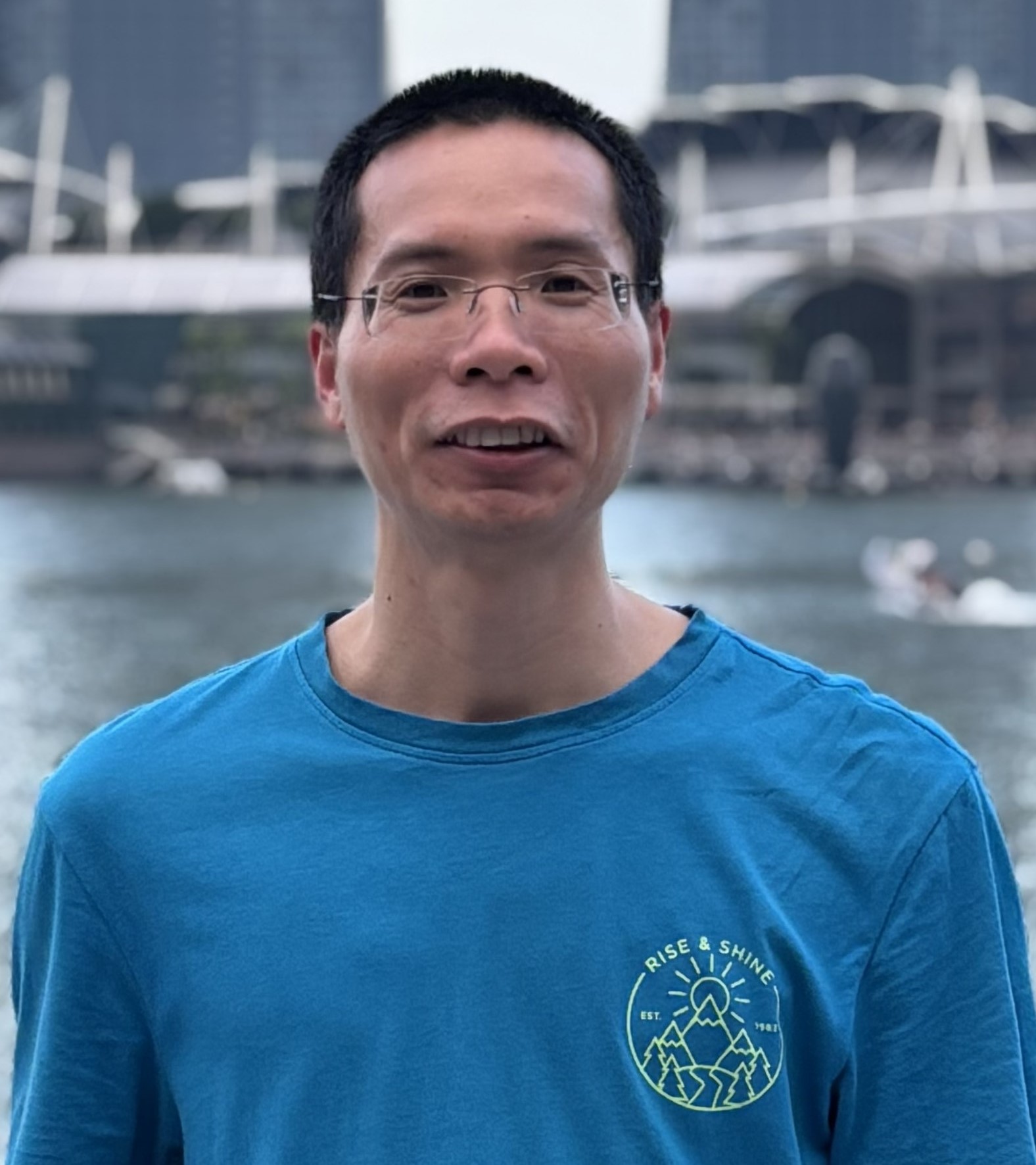}}]
{Jianglin Lan} received the Ph.D. degree from the University of Hull in 2017. He has been a Leverhulme Early Career Fellow and Lecturer at the University of Glasgow since 2022. He was a Visiting Professor at the Robotics Institute, Carnegie Mellon University, in 2023. From 2017 to 2022, he held postdoc positions at Imperial College London, Loughborough University, and University of Sheffield. His research interests include safe AI, fault-tolerant systems, and robotics.
\end{IEEEbiography}

\begin{IEEEbiography}
[{\includegraphics[width=1in,height=1.25in,clip,keepaspectratio]{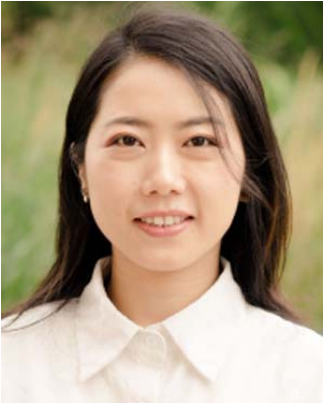}}]
{Lan Zhang} (Member, IEEE) received the B.E. and M.S. degrees from the University of Electronic Science and Technology of China, Chengdu, China, in 2013 and 2016, respectively, and the Ph.D. degree from the University of Florida, Gainesville, FL, USA, in 2020. Since 2024, she has been a tenure-track Assistant Professor with the Department of Electrical and Computer Engineering, Clemson University, Clemson, SC, USA. From 2020 to 2023, she was an Assistant Professor with the Department of Electrical and Computer Engineering, Michigan Technological University, Houghton, MI, USA. Her research interests include wireless communications, distributed machine learning, and cybersecurity for various Internet-of-Things applications.
\end{IEEEbiography}

\begin{IEEEbiography}
[{\includegraphics[width=1in,height=1.25in,clip,keepaspectratio]{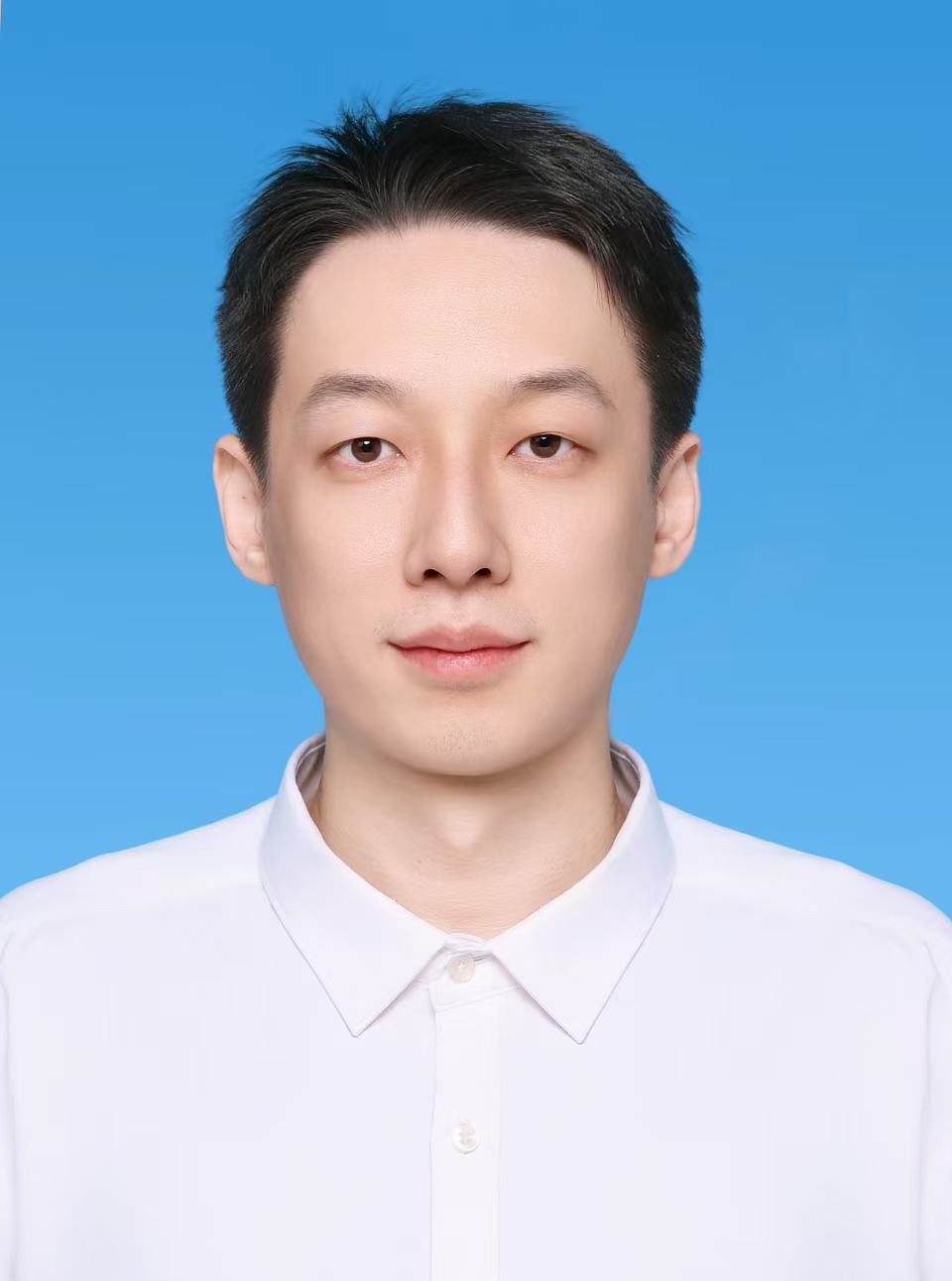}}]
{Xuesong Liu} (Graduate Student Member, IEEE) received the M.Sc. degree in Computer Science from Newcastle University in 2022. He won the Philip Merlin Best Paper Award at Newcastle University in 2022. He is currently pursuing his PhD degree at the University of Glasgow. His research interests include privacy preserving, semantic communication, and information security.
\end{IEEEbiography}

\begin{IEEEbiography}
[{\includegraphics[width=1in,height=1.25in,clip,keepaspectratio]{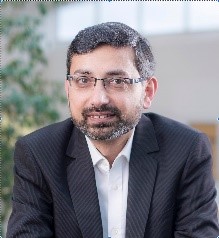}}]
{Muhammad Ali Imran} (Fellow, IEEE) is a Professor of Wireless Communication Systems and Dean of Graduate Studies in College of Science and Engineering. His research interests include self-organized networks, wireless networked control systems, and the wireless sensor systems. He heads the Communications, Sensing and Imaging CSI Hub, University of Glasgow, Glasgow, U.K. He is also an Affiliate Professor with The University of Oklahoma, Norman, OK, USA, and a Visiting Professor with the 5G Innovation Centre, University of Surrey, Guildford, U.K. He has more than 20 years of combined academic and industry experience with several leading roles in multimillion pounds funded projects. His research interests include self-organized networks, wireless networked control systems, and the wireless sensor systems.
\end{IEEEbiography}
\vfill
\end{document}